\documentstyle[aps,prl,eqsecnum,preprint,tighten]{revtex}
\begin{document}
\title {
Quantum field theory of metallic spin glasses}
\author{Subir Sachdev and N. Read}
\address{Department of Physics, P.O. Box 208120, Yale University,
New Haven, CT 06520-8120\\
and\\
Department of Applied Physics, P.O. Box 208284, Yale University,
New Haven, CT 06520-8284}
\author{R. Oppermann}
\address{ Institut f\"{u}r Theoretische Physik, Universit\"{a}t W\"{u}rzburg,
\\
D-97074 W\"{u}rzburg, Germany}
\date{April 4, 1995}
\maketitle

\begin{abstract}
We introduce an effective field theory for the vicinity of a
zero temperature quantum transition between
a metallic spin glass (``spin density glass'') and a metallic quantum
paramagnet.  Following a mean field analysis, we perform a perturbative
renormalization-group study and find that the critical properties
are dominated by static disorder-induced fluctuations, and that
dynamic quantum-mechanical effects are dangerously irrelevant.
A Gaussian fixed point is stable for a finite range of couplings
for spatial dimensionality $d > 8$, but
disorder effects always lead to runaway flows to strong coupling for
$d
\leq 8$.  Scaling hypotheses for a {\em static\/} strong-coupling critical
field theory  are proposed. The non-linear susceptibility has an
anomalously weak singularity at such a critical point. Although motivated
by a perturbative study of  metallic spin glasses, the scaling hypotheses
are more general, and could apply to other quantum spin glass to
paramagnet transitions.
\end{abstract}
\narrowtext

\section{Introduction}
Electronic systems with strong randomness and strong
interactions~\cite{leerama} have been
studied in a
number of experimental systems including doped semiconductors, metallic
alloys, and most
recently in the doped cuprate and doped heavy-fermion compounds. Some of
the most interesting
physics in these materials arises from the complex interplay of the fermionic,
charge-carrying excitations and the spin fluctuations. A number of distinct
equilibrium thermodynamic phases are possible, even at zero temperature
($T$). In the charge
sector, we may have  metallic and insulating phases (within the insulator
we may also
distinguish further between a Mott insulator, with a true $T=0$ charge gap,
and a
Fermi glass, which has
localized, gapless, charged excitations). In the spin sector, the ground
state can either be a spin
glass, in which each spin has an infinite-time memory of its spatially
random moment, or a quantum paramagnet, in which the spin correlations
decay to zero in the
long-time limit. There does not appear to be any fundamental principle
constraining
the relative
positions of the transitions in the charge and spin sectors,
leading to a rich
phenomenology of possible $T=0$ phases and critical points.

Previous work has examined the quantum paramagnet phase
both in the Mott insulator~\cite{bl}
(where the spin fluctuations can be described by a quantum Heisenberg spin
model) and the
metal~\cite{msb}.
Studies of the spin glass phase and its onset have however been
mostly restricted to the insulating phase.
In an infinite-range Heisenberg model of the Mott insulator
an instability of the quantum paramagnet to a possible spin glass was
noted~\cite{Heisenberg}.
Greater progress has been made
in elucidating
the quantum paramagnet to spin glass transition in insulating models of
Ising spins
in a transverse field (which may be appropriate
for insulators with strong crystalline anisotropy) and quantum
rotors~\cite{miller,guo,rsy}. The
methodology of a
recently developed Landau theory for this transition in the Ising and rotor
models~\cite{rsy} will be very
useful to us below.

In this paper, we shall analyze systems in the vicinity of a $T=0$ transition
between a spin glass and a quantum paramagnet occurring while the charge sector
is {\em metallic.}
Our motivation for this study comes partly from recent experiments in
heavy fermion compounds
like $Y_{1-x} U_x Pd_3$~\cite{maple} which appear to show a paramagnet to
spin glass transition
with increasing doping ($x$) in a metallic regime.
However, we shall not make comparisons of our theory with experiments here,
as detailed
studies of the vicinity of the quantum transition are not yet available.

We will begin by introducing in Section~\ref{model}
a quantum field-theory, ${\cal A}$, for metallic spin glasses; our approach
suggests the identification ``spin density glass'' for such systems. In
Section~\ref{mft} we will determine the mean field phase diagram of
${\cal A}$ as a function of a quantum coupling, temperature, and an
external magnetic field.
Fluctuations about mean field will be studied in Section~\ref{fluct}, first
by a perturbative
renormalization group (RG) analysis (Section~\ref{rg}), which finds flows to
strong-coupling for spatial dimensions
$d\leq 8$. These results will then be used (Section~\ref{scaling})
to motivate a scaling scenario for the
strong-coupling region in which the critical fixed point involves only {\em
static\/}
fluctuations induced by the quenched randomness. Dynamic, quantum fluctuations
are dangerously irrelevant at this
static fixed point, and their effects are controlled
by a crossover exponent $-\theta_u \leq 0$. The critical
singularity of the non-linear susceptibility, $\chi_{nl}$, is weakened by a
positive
$\theta_u$, thus a non-divergent, cusp-like, critical singularity in
$\chi_{nl}$ is possible. This scaling scenario generalizes one proposed
earlier for
insulating Ising and rotor spin glasses~\cite{rsy} which had $\theta_u=0$.
Indeed,  there is no fundamental reason why the insulating spin glasses should
not also have
$\theta_u > 0$.

Static, or $\hbar=0$, fixed points have also arisen in studies of some
other  spin systems.
A model
of quantum rotors in a random field was studied some time ago by Boyanovsky and
Cardy~\cite{cb},
and their results can
be interpreted in terms of such a fixed point. However, they
did not realize that the crossovers, and positions of phase boundaries, at
finite $T$ are
modified
by a positive
$\theta_u$; the required modification is related to that discussed by
Weichmann {\em
et. al.\/}~\cite{weichmann} and  Millis~\cite{Millis} in a rather different
physical context, and will also be discussed in this paper.  More recently,
D.~Fisher~\cite{daniel}  has studied the random Ising model in a transverse
field in
$d=1$ and shown that the results are  consistent with a
$\hbar=0$ fixed point: his scaling results however involve an exponential
relationship between energy and length scales, unlike the more usual
power law relationship which we shall find. Finally, in very recent work,
Kirkpatrick and Belitz~\cite{kirkbel} have proposed a scaling scenario for the
metal-insulator transition which appears to have many similarities
to our results below on the metallic spin glass to paramagnet
transition.

\section{Effective Action}
\label{model}
An initial analysis of metallic spin glasses was performed
some time ago by Hertz~\cite{hertz}, although he did not focus on the
vicinity of the $T=0$
quantum  transition. We will study models, similar to those in
Ref~\cite{hertz}, described by
the following class of Hamiltonians:
\begin{equation}
H = -\sum_{i < j , \alpha} t_{ij} c_{i\alpha}^{\dagger} c_{i\alpha}
- \sum_{i < j , \mu} J_{ij}^{\mu} S_{i\mu} S_{j\mu} + H_{int},
\end{equation}
where $c_{i\alpha}$ annihilates an electron on site $i$ with spin
$\alpha=\uparrow, \downarrow$,
and the spin operator $S_{i\mu} \equiv \sum_{\alpha\beta} c_{i\alpha}^{\dagger}
\sigma^{\mu}_{\alpha\beta} c_{i\beta}/2$ with $\sigma^{\mu}$ the Pauli
matrices.
The sites
$i,j$ lie on a $d$-dimensional lattice, the hopping matrix elements $t_{ij}$
are short-ranged and possibly random, and the $J_{ij}^{\mu}$ are Gaussian
random exchange
interactions, possibly with spin anisotropies. The remainder $H_{int}$
includes other
possible short-range interactions between the electrons: we
constrain them so that the ground state of $H$ is metallic.
A version of
$H$ with infinite-range hopping and exchange was studied
recently~\cite{opper} using a static ansatz
for the order parameter but with no additional approximations. We will provide
below a theory which includes dynamic, quantum effects and also applies to
models with finite-range interactions in finite
dimensions.
Our analysis of $H$ will be similar in spirit to the
Stoner model
approaches to the appearance of spin-density-wave order in clean metallic
systems~\cite{hertzold}, except that we now consider
condensation into
a density wave with a random orientation of spins, or a ``spin density glass''.

We now derive a low-energy field-theory for $H$ in the
vicinity of the spin glass to paramagnet transition.
The procedure is
similar to that
of Ref~\cite{rsy}. The metallic nature of the system expresses itself
mainly through
the modification of a single term, which, however, has important consequences.
We begin by defining the spin glass order parameter
\begin{equation}
Q^{ab}_{\mu\nu} (x, \tau_1 , \tau_2 ) = \sum_{i \in {\cal N}(x)} S_{i\mu}^a
(\tau_1 )
S_{i\nu}^b (\tau_2 ) ,
\end{equation}
where $a,b = 1 \ldots n$ are replica indices (the limit $n \rightarrow 0$
is taken at
the end),
$\tau_1, \tau_2$ are Matsubara times, and ${\cal N}(x)$ is a
coarse-graining region around $x$. At $T=0$, the Edwards-Anderson spin glass
order parameter, $q_{EA}$
is the expectation of the replica-diagonal part of $Q$ in
the limit
$|\tau_1-\tau_2| \rightarrow \infty$; however, it is necessary to retain
the time dependence
and all replica components of $Q$ to capture the full structure of the
field theory~\cite{rsy}. The effective
action for $Q$ is obtained by averaging over the $J_{ij}$ randomness in
$H$ after introducing
replicas, decoupling the resulting 8 fermi term by a Hubbard Stratonovich
field $Q$, and integrating out the fermions. This procedure has been
carried out in
Ref~\cite{rsy} for the quantum rotor spin glass, and in Ref~\cite{hertzold}
for spin density wave
formation in metallic systems. It is simple to combine these methods and we
omit
all intermediate steps. The final effective action, ${\cal A}$, is
expressed in terms of a
shifted field $Q \rightarrow Q - C_{\mu\nu}
\delta^{ab} \delta(\tau_1 - \tau_2 )$ where the subtraction only modifies the
uninteresting short time behavior, and the constants $C_{\mu\nu}$ are
chosen so that
the resulting $Q$ is small at low frequencies near the critical
point~\cite{rsy}:
\begin{eqnarray}
&& {\cal A} = \frac{1}{t} \int d^d x \left\{ \frac{1}{\kappa} \int d\tau
\sum_{a\mu}
 r_{\mu} Q^{aa}_{\mu\mu} (x , \tau , \tau ) - \frac{1}{\pi\kappa} \int
d\tau_1 d\tau_2
\sum_{a\mu} \frac{Q^{aa}_{\mu\mu} (x, \tau_1, \tau_2)}{(\tau_1 - \tau_2 )^2}
\right.\nonumber \\
 &&+ \frac{1}{2} \int  d \tau_1
d \tau_2 \sum_{ab\mu\nu} \left[ \nabla Q_{\mu\nu}^{ab} (x, \tau_1, \tau_2 )
\right]^2  - \frac{\kappa}{3} \int  d \tau_1 d \tau_2 d \tau_3
\sum_{abc\mu\nu\rho} Q^{ab}_{\mu\nu} (x, \tau_1 , \tau_2 ) Q^{bc}_{\nu\rho}
(x, \tau_2 , \tau_3 ) Q^{ca}_{\rho\mu}
(x, \tau_3 , \tau_1 ) \nonumber \\
&&~~~~~~~~~~~\left. + \frac{1}{2} \int  d \tau \sum_{a\mu\nu} \left[ u~
Q^{aa}_{\mu\nu} ( x, \tau , \tau) Q^{aa}_{\mu\nu} ( x, \tau , \tau)
+v~ Q^{aa}_{\mu\mu} ( x, \tau , \tau) Q^{aa}_{\nu\nu} ( x, \tau , \tau)
\right]\right\} \nonumber \\
&&~~~~~~~~~~~~~~~~~~ - \frac{1}{2t^2} \int d^d x \int  d \tau_1 d \tau_2
\sum_{ab\mu\nu}
Q_{\mu\mu}^{aa}  (x, \tau_1 , \tau_1 ) Q_{\nu\nu}^{bb} ( x, \tau_2 ,
\tau_2 ) ~~~+~~~ \cdots~.
\label{landau}
\end{eqnarray}
We have only displayed the small subset of terms which will be
important near the
critical point. We have allowed a $\mu$ dependence in $r_{\mu}$ to reflect
possible spin
anisotropies; the less-important $\mu$ dependence of other couplings
has been suppressed.
The metallic nature of the system is reflected in
the second term which has a long-range $1/(\tau_1 - \tau_2)^2$ interaction
in time;
the power-law decay is a consequence of the gapless particle-hole
spin excitations which lead
to the dependence
\begin{equation}
\left[ \left\langle S_{i\mu} (\tau_1) S_{i \nu} ( \tau_2 ) \right\rangle
\right]
\sim \delta_{\mu\nu} (\tau_1 - \tau_2 )^{-2}
\end{equation}
in any metallic paramagnet~\cite{leerama,hertzold} (the angular brackets
represent
an average over
quantum and thermal fluctuations, and the square brackets are an average
over randomness).
This behavior is of course only valid at large $|\tau_1 - \tau_2|$
and we cut it off at short time differences such that its Fourier transform
is $\sim -|\omega|$ for small $\omega$.
The remaining terms in ${\cal A}$ are identical to those obtained in
Ref~\cite{rsy}
for the rotor model.
These are the most general terms, of low order in $Q$, which are
local in space-time
and consistent with underlying symmetries. In particular, the time
arguments of $Q$
associated with different replica indices must always be integrated
independently because the
disorder is static. ``Quantum-mechanical'' interactions occur only within the
same replica, allowing a gradient expansion about
the equal-time point for such terms.
A more detailed discussion of these criteria can be found in Ref~\cite{rsy}.
The particle-hole continuum will induce non-local corrections to these
terms, but none are as
important as that in the term linear in $Q$.
The symmetries also allow a `mass' term $\sim \left( Q_{\mu\nu}^{ab} (x,
\tau_1 , \tau_2 )
\right)^2$, but such a term is redundant as it can be removed by the shift
$Q \rightarrow Q - C_{\mu\nu}
\delta^{ab} \delta(\tau_1 - \tau_2 )$; the shift has a delta function in
time and thus does not
modify the long-time, low frequency behavior that we are interested in.

We now discuss the physical significance of the couplings in
${\cal A}$; for more details the reader is referred again to Ref~\cite{rsy}
and we highlight only the main points here.
First note that there are more terms than coupling constants, but it is
easy to check that
rescalings of $x$ and $\tau$ always allow one to reach the form
chosen.
The coupling
$r_{\mu}$ multiplies what turns out to be ``thermal operator'' which tunes the
system across the transition.
The important non-linearity is the cubic $\kappa$ coupling which is induced
by disorder effects
and involves no exchange of energy between the $Q$ fields; there is a
$1/\kappa$ in the linear
term to ensure that the bare $Q$ propagator is independent of $\kappa$.
The only terms involving
energy exchange are the quadratic
$u$ and $v$ terms which represent the ``quantum-mechanical'' interactions
between the fermions.
Finally, the $1/t^2$ term in ${\cal A}$ represents
disorder fluctuation
effects and arises from fluctuations in the local
position of the critical
point as determined by $r_{\mu}$.

\section{mean field theory}
\label{mft}

We now consider the mean field (or tree-level) properties of ${\cal A}$.
We will only consider two extreme
limits of the spin
anisotropy---Ising-like, when $r_1 \ll r_2 , r_3$, and Heisenberg-like,
when $r_1 = r_2 = r_3$.
We will drop the vector $\mu$ index except where necessary, and represent
the effective number of components by
$M$ ($=1$ for the Ising case,
and $=3$ for the Heisenberg case).

In Ref~\cite{rsy} we found that, as in the classical spin glass,
a theory with only a
cubic non-linearity was replica symmetric, even in the spin glass phase;
replica
symmetry breaking appeared only upon including a certain quartic coupling
which was
formally irrelevant at the quantum critical point. Much of this structure
carries
over unchanged to the metallic spin glass, and we shall not dwell on it
here. We will
restrict our considerations to the replica symmetric theory which contains
all of
the dynamic effects which lead to the important differences between metallic
and
insulating spin glasses. We therefore make the following
replica-symmetric, $x$-independent ansatz for the mean field value of $Q$:
in Matsubara
frequencies (which are integral multiples of $2 \pi k_B T / \hbar $ as usual)
we write
\begin{equation}
Q^{ab} ( \omega_1 , \omega_2 )
= \beta^2  \delta_{\omega_1,0} \delta_{\omega_2 , 0} q +   \beta
\delta^{ab} \delta_{\omega_1 +
\omega_2 , 0} \chi (i \omega_1 ),
\label{ansatz}
\end{equation}
where $\beta=\hbar /k_B T$, and $\chi$ is the local dynamic spin susceptibility
(we will absorb a factor of $k_B / \hbar$ into
$T$ from here
on).
The first term is the spin glass order
parameter---note that it is independent of replica indices, and, therefore, the
replica diagonal and off-diagonal components of $q$ are equal and $q=q_{EA}$.
In a theory with only a cubic
non-linearity (to which we shall restrict ourselves here)
the equality of all the replica components of $q$ holds at all $T$;
upon including
higher-order terms in
${\cal A}$, the equality between all the replica components persists at
$T=0$, but there are thermal corrections at any
non-zero $T$ which distinguish the diagonal and off-diagonal
components and which
also break replica symmetry~\cite{rsy}.

 We now insert (\ref{ansatz}) into
${\cal A}$ and determine
the saddle point. This determines the spin glass to paramagnet
phase boundary at
$r=r_c (T)$ and the order parameter
\begin{equation}
q = \left\{ \begin{array}{cc}
(r_c (T) - r)/(\kappa (u + Mv) & \mbox{for $r < r_c
(T)$} \\
0 & \mbox{otherwise}
\end{array} \right. ,
\end{equation}
with
\begin{equation}
r_c (T) = r_c - c (u + Mv ) T^{3/2},
\label{rcres}
\end{equation}
where $c = \sqrt{\pi/2} \zeta(3/2)$, $r_c \equiv r_c (0) \sim
\Lambda_{\omega}^{3/2}$
is dependent on the large frequency cutoff
$\Lambda_{\omega}$ (see
Fig~\ref{phasediag}). The local, dynamic spin susceptibility has imaginary part
\begin{equation}
\chi^{\prime\prime} ( \omega ) = -\frac{1}{\kappa} \mbox{Im}
\sqrt{-i\omega + \Delta}
= \frac{1}{\sqrt{2} \kappa} \frac{\omega}{\sqrt{\Delta + \sqrt{\omega^2 +
\Delta^2}}}.
\end{equation}
Note that the spin fluctuations are gapless, and the crossover from
paramagnetic to
critical fluctuations occurs at a
frequency scale $\Delta$; $\Delta$ also determines the correlation length
$\xi \sim
\Delta^{-1/4}$ that appears in spatial correlation functions, which can be
found as
Gaussian fluctuations around the saddle point. The value of $\Delta$ has a
rather complicated dependence on
$T$ and $r$ and we describe
its limiting behavior in the five different regimes of
Fig~\ref{phasediag}---there are smooth
crossovers between these regimes.
Within the spin glass phase, we have $\Delta=0$, $\chi^{\prime\prime} \sim
\mbox{sgn} ( \omega ) \omega^{1/2}$ everywhere and there
are no crossovers in the present approximation. However, we expect that
there will be
a crossover between a region characterized by the quantum ground state (I)
to a region
dominated by thermal, critical fluctuations (II) and such a crossover
boundary has been
shown in Fig~\ref{phasediag}. In the paramagnetic phase the present
approximation is
much richer, and shows all the expected crossovers. The scale $\Delta$ is
determined by the equation
\begin{equation}
\Delta = r - (u+Mv) \frac{1}{\beta} \sum_{\omega} ( |\omega| + \Delta)^{1/2} .
\label{deltaeqn}
\end{equation}
Solution of (\ref{deltaeqn}) yields four different regimes
of behavior (II-V) which we describe in turn. \\
(II) $|r-r_c (T)| \ll (u+Mv)^2 T^2$: this is closest to the phase
boundary, and is the region with classical fluctuations.
We have
$\Delta = ((r-r_c (T))/(T(u+Mv)))^2$---note that $\Delta$ depends on the
{\em square} of the distance $r-r_c (T)$ from the transition (as shown in
Ref~\cite{rsy}, this is crucial for obtaining the classical exponent
$\nu=1/2$). \\
(III) $(|r - r_c |/(u+Mv))^{2/3} \ll T$: this
is the `quantum-critical' region in that
$T$ is the most significant energy scale and the system behaves as if it is
at the critical coupling
$r=r_c$. Here we find, to lowest order in $u,v$ that $\Delta = c (u + M v)
 T^{3/2}$. Note that the expected quantum-critical scaling $\Delta \sim
T$~\cite{jinwu}
is violated. This is a consequence of the fact that all $T$ dependent
corrections are controlled by quantum interactions $u,v$ which are
irrelevant at
the critical point (see below)---in other words $\theta_u > 0$
spoils the naive
$\Delta \sim T$ scaling. A similar interpretation can be given to the
position of the phase
boundary $r_c (T)$. \\
(IV) $r-r_c \ll T \ll ((r-r_c )/(u+Mv))^{2/3}$ {\em and\/} (V) $T \ll r - r_c$:
these are the
`quantum-disordered' regions in that
$T$ dependent corrections are secondary and to leading order in $T,u,v$ we have
$\Delta = r-r_c$. The subleading terms in $\Delta$ are different in the two
regimes:
in regime IV, $\Delta (T) - \Delta (0) =
c (u + M v) T^{3/2}$, while in regime V,
$\Delta(T) - \Delta (0) = (u+Mv) \pi T^2 /(6 \sqrt{r-r_c})$. This
subdivision of the
quantum-disordered region is similar to that found in a different context
in Ref~\cite{Millis},
and is also a consequence of the dangerous irrelevancy of $u,v$.

All of the above crossover boundaries and exponents are of course
characteristics of the
mean field theory, which can, in general, be modified by fluctuations---we
will indicate
the nature of these modifications in the discussion in Section~\ref{scaling}.

\subsection{Phases in a magnetic field}
\begin{figure}
\caption{$T=0$ phase diagram of ${\cal A}$ for the Heisenberg case, in a field
$H$. $H_{GT}$ is the Gabay-Toulouse boundary, $q_L
= \left[\langle S_{\mu} \rangle^2 \right]$ for $\mu$ along the field direction,
and similarly  $q_T
= \left[\langle S_{\mu} \rangle^2 \right]$ for $\mu$ perpendicular to the
field.
}
\label{gt}
\end{figure}
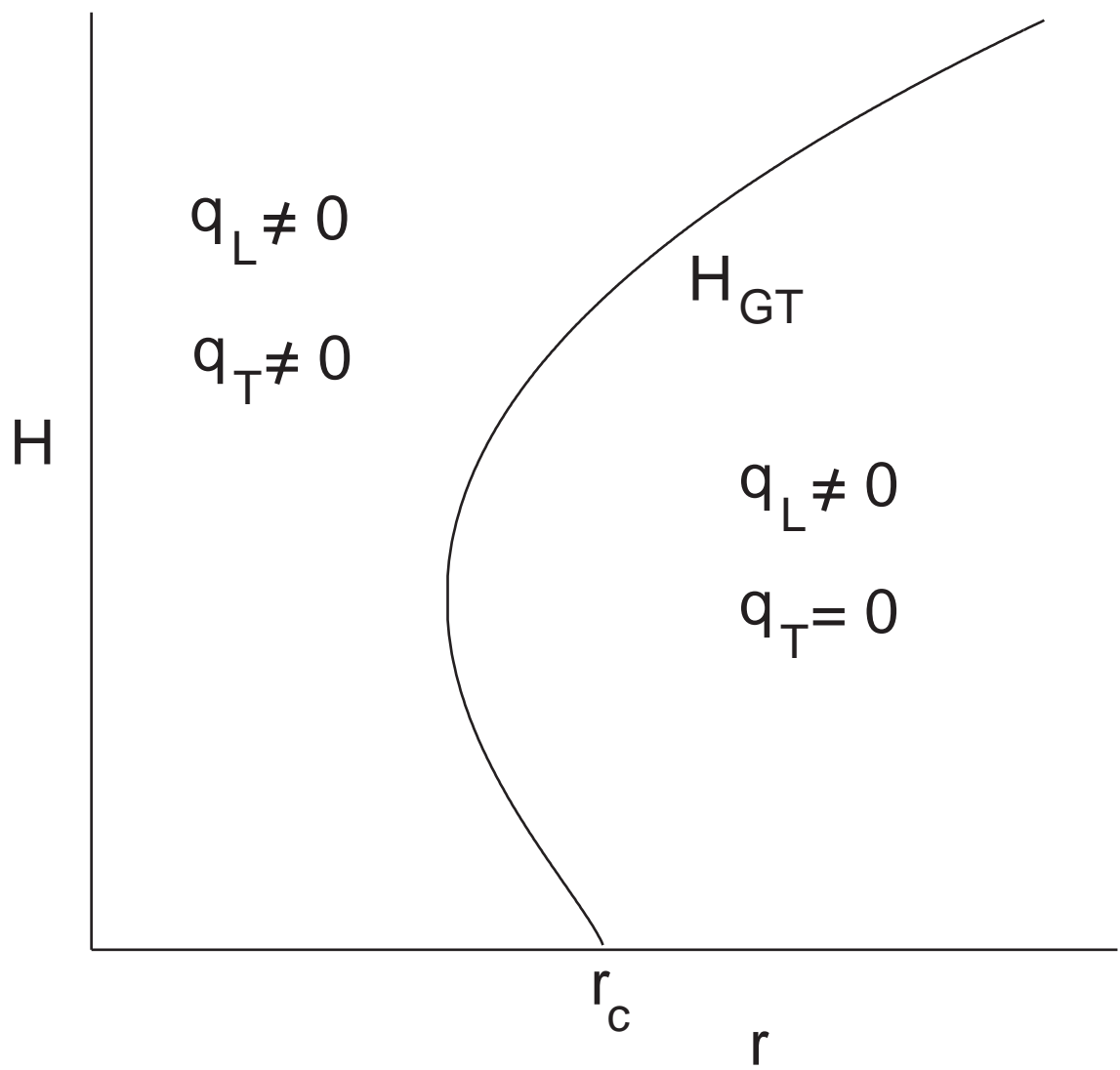

We complete our discussion of mean field theory by discussing the effect of an
external magnetic field, $H_{\mu}$ on ${\cal A}$. The additional terms induced
by $H$ can be determined following Refs~\cite{im} which examined the
effects of $H$ on spin-density wave formation in clean systems---by this
method we
found
\begin{eqnarray}
{\cal A} \rightarrow {\cal A} - && \frac{g}{t}  \int d^d x d\tau_1 d
\tau_2
\sum_{ab} Q^{ab}_{\mu\nu} (x, \tau_1 , \tau_2 ) H_{\mu} H_{\nu} \nonumber \\
- && \frac{1}{t} \int d^d x d \tau \sum_{a} \left( i \alpha_1
\epsilon_{\mu\nu\lambda}
H_{\lambda} \left. \frac{\partial
Q^{aa}_{\mu\nu} (x, \tau_1 , \tau_2 ) }{\partial\tau_1}  \right|_{\tau_1 =
\tau_2 =
\tau} + \alpha_2 H_{\lambda} H_{\lambda} Q^{aa}_{\mu\mu} (x, \tau, \tau)
\right.\nonumber\\
&&~~~~~~~~~~~~~~~~~~~~~~~~~~~
 + \alpha_3 (H_{\mu} H_{\nu} -
 H_{\lambda} H_{\lambda} \delta_{\mu\nu}) Q^{aa}_{\mu\nu} (x,
\tau, \tau)
\Biggr) .
\label{field}
\end{eqnarray}
The field has several different competing effects. The first term,
proportional to the coupling $g$, is the
static
paramagnetic susceptibility of the fermions which polarizes the spins along the
field, and which always dominates at small $H$. The $\alpha$ terms
account for the precession of the spins in the plane perpendicular to $\vec{H}$
and the energetic contribution of quantum fluctuations about the static spin
directions: as in clean antiferromagnets~\cite{conserve} we expect these
terms to
prefer magnetic order in a plane perpendicular to $H$ (and so $\alpha_3 > 0$).
We now consider some cases separately: \\
({\em i\/})
{\em Ising spins, $H$ along easy-axis\/}---only the term proportional to $g$ in
(\ref{field}) need be
considered as the $\alpha$ terms are never important.
The finite field phase diagram, the field dependence of observables, and
the position of
the Almeida-Thouless boundary~\cite{book}, are essentially identical
to that for the insulating Ising model considered earlier~\cite{rsy},
and will therefore not be considered here. The only difference in the metallic
case is that no logarithms are present---{\em e.g.\/} the free energy at
$r=r_c$
and
$T=0$ that depended on $H$ as $H^{8/3} /\log^{1/3} H$in Ref~\cite{rsy}, here
varies as
$H^{8/3}$.\\ ({\em ii\/}) {\em Ising spins, $H$ perpendicular to
easy-axis\/}---now the $g$
term in (\ref{field}) couples only to non-critical
components
of $Q$ and is not important; after integrating out the non-critical $Q$
 we find that the main consequence
of the
$\alpha$ terms
is to induce a shift $r_1 \rightarrow r_1 + \alpha^{\prime} H^2$ in
the position of the
critical point. \\
({\em iii\/}) {\em Heisenberg spins\/}---in finite field we
now have to
allow for the possibility of spin glass order appearing in the plane
perpendicular to $H$; the onset of this order is the Gabay-Toulouse
transition at
$H=H_{GT}$~\cite{book}.
Let us take a field $H$ pointing along the $\mu=3$ direction. The
subsequent mean
field theory is most convenient in a circularly-polarized basis for the vector
components of $Q$:
we take $Q_{33} = Q_L$, $Q_{11} = Q_{22} = (Q_{+-} + Q_{-+})/2$,
$Q_{12} = - Q_{21} = i(Q_{+-} - Q_{-+})/2$, and all other vector components
of $Q=0$.
We make the same ansatz as in (\ref{ansatz}) for the frequency and replica
dependence
of $Q_L$, $Q_{+-}$ and $Q_{-+}$ by introducing the quantities $q_L$,
$\chi_L$, $q_{+-}$
{\em etc.\/}. It is then not difficult to solve the resulting mean field
equations.
It is slightly more convenient to approach the Gabay-Toulouse boundary from the
Gabay-Toulouse phase with spin glass order in the transverse direction: $q_{+-}
= q_{-+} \neq 0$. The solution of the mean field equations for this case are
\begin{eqnarray}
q_{+-} &=& q_{-+}=q_T \nonumber \\
\chi_{+-} (i \omega) &=& \chi_{-+}^{\ast} (i \omega) = - \frac{1}{\kappa}
( |\omega| + i \alpha_1 H \omega )^{1/2} \nonumber \\
q_{L} &=& \frac{g H^2}{4 \sqrt{\Delta}} \nonumber \\
\chi_L (i \omega ) &=& - \frac{1}{\kappa} ( |\omega| + \Delta )^{1/2}~,
\end{eqnarray}
where the Gabay Toulouse spin glass order parameter $q_T$ and the frequency
scale $\Delta$
are determined by the solutions of the two equations
\begin{eqnarray}
\Delta &=& r + \alpha_2 H^2 + (u+v) \left( \frac{\kappa g H^2}{4
\sqrt{\Delta}} -
\frac{1}{\beta} \sum_{\omega} (
|\omega| +
\Delta )^{1/2} \right) +  2v \left( \kappa q_T - \frac{1}{\beta} \sum_{\omega}
(|\omega|+ i \alpha_1 H \omega )^{1/2} \right) \nonumber \\
0 &=& r + (\alpha_2 -\alpha_3 ) H^2 + v \left( \frac{\kappa g H^2}{4
\sqrt{\Delta}} -
\frac{1}{\beta} \sum_{\omega} (
|\omega| +
\Delta )^{1/2} \right)  \nonumber \\
&~&~~~~~~~~~~~~~~~~~~~~~~~~~~~~~~~~~~~~~~~~~~~
+(u + 2 v) \left( \kappa q_T - \frac{1}{\beta} \sum_{\omega}
(|\omega|+ i \alpha_1 H \omega )^{1/2}\right) .
\end{eqnarray}
The Gabay-Toulouse boundary is determined by imposing the condition $q_T=0$
on these two equations, which gives us a line $H= H_{GT} (r)$ in the $r-H$
plane.
The result of such a computation at $T=0$, is shown in
Fig~\ref{gt}.
For small $H$, the first term in (\ref{field}) dominates and we find
$H_{GT} \sim (r_c
- r)^{3/4}$. For large $H$, the $\alpha$ terms take over,
and for $\alpha_3 > 0$
 we find that $H_{GT}$ turns over and extends to $r>r_c$
as $H_{GT} \sim (r-r_c )^{1/2}$ (see Fig~\ref{gt}); for sufficiently
negative $\alpha_3$
this turn over will not occur.

\section{Fluctuations}
\label{fluct}
We begin by a perturbative RG analysis of fluctuations, which will
unfortunately
not be of much direct utility as there is runaway flow to strong coupling below
$d=8$. Nevertheless, the structure of this analysis will help motivate a
general scaling scenario which we will describe subsequently.

\subsection{Perturbative RG}
\label{rg}
The perturbative RG analysis is quite similar to that of Ref~\cite{rsy}. The
main difference at tree level will be that the dynamic exponent $z$ is $z=4$
rather than $z=2$. This difference has the important consequence of
now making the $u,v$ couplings dangerously irrelevant, which in turn leads to
a positive $\theta_u$.

The RG begins with the rescalings
\begin{equation}
x^{\prime} = x/s~~~,~~~ \tau^{\prime}
= \tau/s^z~~~,~~~
t^{\prime} = t s^{-\theta}~~~,~~~Q^{\prime} = Q
s^{(d-\theta+2z-2+\eta)/2}.
\label{rgbasic}
\end{equation}
The exponents
$z, \eta$ have their usual meaning, while $\theta$ is introduced to allow
for violations of
hyperscaling: we will have $\theta > 0$ causing $t$ to flow to 0, and behave
as a dangerously
irrelevant variable.
The irrelevant coupling $t$ and its exponent $\theta$, should not
be confused with the couplings controlling quantum-mechanical effects
and the exponent $\theta_u$ which will be discussed momentarily.
At tree level (or equivalently, at the gaussian fixed point),
the above
rescalings leave
${\cal A}$ invariant provided we modify the couplings
\begin{eqnarray}
r^{\prime} = r s^{z}~~~~~~&&~~~~~~~\kappa^{\prime} = \kappa
s^{(6+\theta-d-3\eta)/2}\nonumber \\
u^{\prime} = u s^{2-z-\eta}~~~~~~&&~~~~~~~v^{\prime} = v s^{2-z-\eta} .
\label{rescalings}
\end{eqnarray}
and choose the exponents
\begin{equation}
z = 4~~~~~~~~~~~~~~\eta=0~~~~~~~~~~~~~~~\theta=2 .
\end{equation}
Thus the cubic non-linearity $\kappa$
becomes relevant for $d$ below $8$, and the rescaling of
$r$ gives us the gaussian exponent $\nu=1/4$; note the correlation length
is given by
$\xi \sim \Delta^{-1/4}$ in the notation of Section~\ref{mft}.
The most important point, and the key difference from Ref~\cite{rsy},
is that $u$ and $v$ are now irrelevant with exponent $-2$.
As these are the only couplings associated with quantum effects,
we introduce a new crossover exponent, $-\theta_u$ which will control
the `dangerously irrelevant' consequences of quantum fluctuations. At tree
level we clearly have $\theta_u = 2$. Finally, note that for $d$ above 8,
$\kappa$
also becomes dangerously irrelevant about the Gaussian fixed point.

It is straightforward to extend the above analysis to include one-loop
diagrams.
Using the diagrams discussed in Ref~\cite{rsy}, we find ($s=e^{\ell}$)
\begin{equation}
z(\ell) =4 + 8 \kappa^2 (\ell )~~~,~~~\eta=2\kappa^2 (\ell )~~~,~~~\theta=2,
\end{equation}
and the flow equations
\begin{equation}
\frac{dr ( \ell )}{d\ell} = z r (\ell ) - a \kappa^2 (\ell
)~~~;~~~\frac{d\kappa
(\ell ) }{d\ell} =
\frac{8-d}{2} \kappa (\ell ) + 9 \kappa^3 (\ell ).
\label{rgeqn}
\end{equation}
We have absorbed various phase-space factors into the couplings
(see~\cite{rsy}),
and
$a$ is an uninteresting
positive constant.
The irrelevant couplings $u,v$ were set equal to 0 at the outset.
There is no stable fixed point of (\ref{rgeqn}) for real $\kappa$ below $d=8$.
Above $d=8$, the Gaussian fixed point is stable, but its domain of attraction
is limited to a region which vanishes as $d$ approaches 8 from above.
For $d \leq 8$, and for
all physical initial conditions,
the coupling $\kappa$ flows to strong coupling, making quantitative
computation of exponents impossible in the present approach.

\subsection{Scaling Hypotheses}
\label{scaling}
We will now discuss a non-perturbative scaling scenario for quantum spin
glasses,
assuming that the structure of the dangerously irrelevant variables remains
similar to
that found in the perturbative analysis above. We will consider a {\em static}
strong-coupling critical theory with two dangerously irrelevant directions:
one associated with a coupling analogous to $t$ which controls disorder
fluctuation effects
and has exponent $-\theta$, and a second associated with dynamic, quantum
mechanical
effects (couplings $u,v$) and exponent $-\theta_u$. In the previous analysis
of insulating spin glasses~\cite{rsy} only $t$, the first of these dangerously
irrelevant couplings, was present; our present scaling relations reduce to the
earlier ones upon putting $\theta_u=0$. As we shall see below, a positive
$\theta_u$
has new and important physical consequences. Although the analysis below is
clearly motivated by our mean field theory of metallic spin glasses above,
there is no
fundamental reason why the insulating models considered in Ref~\cite{rsy}
should not also
have $\theta_u > 0$.

It is helpful to discuss non-perturbative effects by considering the
scaling behavior
of observable correlation functions. Among two-point correlators of $Q$,
there are
three independent observables~\cite{rsy}: in the paramagnetic phase these
are the spin glass
susceptibility,
$G$,
\begin{eqnarray}
G (x-y, \tau_1 - \tau_2 , \tau_3 - \tau_4 )
&\equiv&  \sum_{\mu\nu}
\left[ \left\langle S_{i\mu} (\tau_1 ) S_{j\mu} (\tau_2 ) \right\rangle
\left\langle S_{i\nu} (\tau_3 ) S_{j\nu} (\tau_4 ) \right\rangle \right]
\nonumber \\
&=&  \lim_{n \rightarrow 0}
\frac{1}{n(n-1)} \sum_{a\neq b,\mu\nu} \left\langle\left\langle Q^{ab}_{\mu\nu}
(x, \tau_1 , \tau_3 ) Q^{ab}_{\mu\nu} ( y, \tau_2 , \tau_4 )
\right\rangle\right\rangle
\label{defgd2}
\end{eqnarray}
(the double angular brackets represent averages with a replicated,
translationally invariant
action like ${\cal A}$);
the quantum mechanically disconnected correlator, $G^d$,
\begin{eqnarray}
G^{d} ( i-j, \tau_1 - \tau_2 , \tau_3 - \tau_4 ) &\equiv&
\left[ \left\langle S_{i\mu} (\tau_1 ) S_{i\mu} (\tau_2 ) \right\rangle
\left\langle S_{j\nu} (\tau_3 ) S_{j\nu} (\tau_4 ) \right\rangle \right] -
\mbox{subtractions}
\nonumber \\
&=& \lim_{n \rightarrow 0}
\frac{1}{n(n-1)} \sum_{a\neq b} \left\langle\left\langle Q^{aa}_{\mu\mu}
(x, \tau_1 , \tau_2 ) Q^{bb}_{\nu\nu} ( y, \tau_3 , \tau_4 )
\right\rangle\right\rangle -\cdots;
\end{eqnarray}
and the connected correlation function, $G^c$,
\begin{eqnarray}
G^c_{\mu\nu\rho\sigma}
(i - j , \tau_1 - \tau_4 , \tau_2 - \tau_4 , \tau_3 -
\tau_4) &\equiv&  \left[\left\langle S_{i\mu} ( \tau_1 ) S_{i\nu} (\tau_2 )
S_{j \rho} ( \tau_3 ) S_{j \sigma} ( \tau_4 ) \right\rangle\right]
-\mbox{subtractions}
\nonumber \\
&=& \lim_{n \rightarrow 0}
\frac{1}{n} \sum_{a} \left\langle\left\langle Q^{aa}_{\mu\nu}
(x, \tau_1 , \tau_2 ) Q^{aa}_{\rho\sigma} ( y, \tau_3 , \tau_4 )
\right\rangle\right\rangle
-\cdots .
\label{defgc}
\end{eqnarray}
The non-linear susceptibility, $\chi_{nl}$, is given by the integral over
space and time
of $G^c$.
The correlator $G^c$ is non-zero only because of the ``quantum interactions''
$u,v$ , and therefore carries a prefactor of $u,v$; in contrast $G$ and
$G^d$ are
non-zero even in a purely static theory.  Under the rescalings
(\ref{rgbasic}) we
may conclude from arguments similar to those in Ref~\cite{rsy} that $G^d$
and $G$
scale as
\begin{eqnarray}
G^d (x, \tau, \tau) &\sim& x^{-(d+2z-\theta-2+\eta)} \label{gdeq} \\
G (x, \tau, \tau) &\sim& x^{-(d+2z-2+\eta)} \label{geq}~,
\end{eqnarray}
for fixed $\tau/x^z$ at criticality.
The scaling dimensions of $G$ and $G^d$ differ because $G$ carries a
prefactor of the
dangerously irrelevant variable $t$, while $G^d$ does not. Finally, $G^c$
carries
a prefactor of $t$, and an additional prefactor of the irrelevant quantum
interactions:
hence
\begin{equation}
G^c (x, \tau, \tau, \tau) \sim x^{-(d+2z+\theta_u-2+\eta)}.
\label{gceq}
\end{equation}
Indeed, one can consider the 3 equations (\ref{gdeq})-(\ref{gceq}) as the
definition
of the three independent exponents $\eta$, $\theta$, and $\theta_u$.
In the previous analysis~\cite{rsy}, $\theta_u=0$ and hence $G$ and $G^c$
had the same scaling dimension.
By taking the spacetime integral of (\ref{defgc}), we can deduce that the
non-linear
susceptibility $\chi_{nl}$ behaves as
\begin{equation}
\chi_{nl} \sim |r-r_c|^{-(2-\eta-\theta_u)\nu}
\end{equation}
near the $T=0$ quantum critical point. For sufficiently large $\theta_u$,
$\chi_{nl}$ need not diverge.

The existence of a static critical theory, and the associated
positivity of $\theta_u$ ,
has important consequences for the finite $T$ behavior away from the
critical point.
Recall that $T$ only appears as a finite-size length, $1/T$, along the time
direction, and
hence scaling~\cite{cb} implies that its scaling
dimension is $z$.
However some power of a combination of interactions like $u,v$ must appear in
any frequency scale and hence we expect that the naive scaling of finite
$\omega$ correlators as functions of
$\omega /  T$~\cite{jinwu} will now be modified. A related
modification of naive scaling
has been discussed in Refs~\cite{weichmann,Millis} for some clean systems,
and we will now
present a similar analysis. It is useful to consider a
simple model of the renormalization group flows near the quantum-critical
point at low
temperatures. Let us move away from the quantum critical point ($r=r_c$, $T=0$)
by perturbing the system along
the single, relevant eigendirection by the amount $r-r_c$, and along the
least irrelevant
eigendirection which involves terms with frequency exchange by the amount $u$.
For small $r$, $u$, and $T$ we expect flow equations like
\begin{eqnarray}
\frac{dT(\ell )}{d\ell} &=& z T(\ell ) \nonumber \\
\frac{dr(\ell )}{d\ell} &=& \frac{1}{\nu} (r(\ell )-r_c) + u f(T(\ell ))
\nonumber \\
\frac{du(\ell )}{d\ell} &=& -\theta_u u(\ell )~,
\label{uflow}
\end{eqnarray}
where $f(T)$ ($f(0)=0$) is some function arising from thermal occupation of the
short distance modes of the order parameter fluctuations which are begin
integrated out. The key property of (\ref{uflow}) is that a $T$ dependence is
induced into the flow of the relevant coupling $r$ only via the irrelevant
coupling
$u$.  The integral of (\ref{uflow}) is
\begin{equation}
r(\ell) - r_c = (r -r_c) e^{\ell/\nu} + u e^{\ell/\nu} \int_0^{\ell} d
\ell^{\prime} e^{-
(\theta_u + 1/\nu)
\ell^{\prime}} f( T e^{z \ell^{\prime}})
\end{equation}
(as is customary, we have abbreviated  $r(0)=r$, $u(0)=u$ and $T(0)=T$) .
We now change integration variables to $\zeta = T e^{z \ell^{\prime}}$, and
integrate
to the correlation length $\xi = e^{\ell = \ell^{\ast}}$ at which
$r(\ell^{\ast} ) - r_c =
1$ to obtain
\begin{equation}
1 = \xi^{1/\nu} \left[ r - r_c + \frac{u T^{(\theta_u + 1/\nu)/z}}{z}
\int_T^{T \xi^z} \frac{d\zeta}{\zeta} \zeta^{-(\theta_u+1/\nu)/z} f(\zeta )
\right].
\label{zeta}
\end{equation}
It is now possible to deduce scaling properties provided it is permissible
in the critical
region to set the lower and upper limits of the integral in (\ref{zeta}) to
zero
and infinity respectively. As $f(T)$
represents thermal contribution of short distance modes we expect it to vanish
as $T \rightarrow 0$; these modes however do mix with the particle-hole
continuum of the metal,
leading us to expect a linear density of states at low energies
even at short distances, and therefore
$f(T)\sim
T^2$ for small $T$. In the opposite large $T$ limit, all modes must become
classical,
and therefore $f(T) \sim T$. For these asymptotic behaviors in $f(T)$, the
limits on the
integration can be extended provided $z \nu < 1+\theta_u \nu < 2 z\nu$. We
then obtain
at  $r = r_c$ but $T$ finite $\xi^{-1} \sim u^{\nu} T^{(1 + \theta_u \nu)/z}$.
In this same region, a similar reasoning implies that
 the local dynamic spin susceptibility
will scale as
\begin{equation}
\chi^{\prime\prime} ( \omega ) = \omega^{(d-\theta-2+\eta)/2z} \phi\left(
 \frac{ K \omega}{u ^{z\nu} T^{1+ \theta_u \nu}} \right),
\end{equation}
for some universal scaling function $\phi$, and non-universal constant $K$.
At tree-level this gives us a frequency scale $\sim T^{3/2}$ which is
consistent
with the results of Section~\ref{mft}. Similarly, the position of the finite
temperature spin glass to paramagnet boundary (at $r=r_c (T)$) will scale near
the quantum critical point at $r=r_c$ and $T=0$ as
\begin{equation}
r_c - r_c (T)  \sim u T^{(1 + \theta_u \nu)/z\nu} ;
\end{equation}
Again, this agrees with the tree-level result (\ref{rcres}).
A very similar result applies to the boundary between regions III and IV
of Fig~\ref{phasediag} which occurs at $r-r_c
\sim u T^{(1 + \theta_u \nu)/z\nu}$, while the boundary between regions IV and
V is at $r-r_c \sim T^{1/z\nu}$.

All of the results discussed so far in this section have been obtained
using only rather
general scaling ideas. In particular, they do not rely on the particular
form of the action
${\cal A}$. We will now obtain a few results which {\em do\/} rely on
explicit features
of ${\cal A}$, and their validity is therefore somewhat more questionable.

A simple argument can be given to fix the value of $z$,
using the manner in which time dependence enters into ${\cal A}$.
 Consider
a correlator of the $Q$ fields in which all external frequencies have been
fixed at
the same frequency $\omega$. As
the critical field theory is static, and because the $Q$ field is bilocal
in time,
$\omega$ will act simply as an external source which shifts the value of the
``thermal'' coupling
$r \rightarrow r + |\omega|$, as is apparent from the first two terms in
${\cal A}$;
for insulating Ising and rotor models
the corresponding shift is
$r \rightarrow r + \omega^2$. As the scaling dimension of $r$ is $1/\nu$,
this gives
us the scaling relation
\begin{equation}
z \nu = \left\{ \begin{array}{cc}
1 & \mbox{metallic spin glasses} \\
1/2 & \mbox{insulating Ising and rotor spin glasses}
\end{array} \right.~.
\label{znu}
\end{equation}
We emphasize that both results rely on the assumption of a static critical
theory;
this assumption was not made in the analysis of Ref~\cite{rsy}.
We also note in passing that the present argument fixing the value of $z$
cannot be applied to the random-field
quantum rotor model of Ref~\cite{cb} (which also had a static critical point),
because the
same  external frequency $\omega$ does not flow through all internal
propagators in
this case, and some propagators are always at zero frequency.

We now ask whether there is
a {\em classical\/} statistical mechanics field theory which is also described
by the static critical point postulated above.
We are only able to answer this question within the confines of
perturbation theory:
a perturbative
expansion in $\kappa$ suggests that the relevant field theory is that
describing
singularities along the imaginary field, $ih$, axis in a $d$-dimensional
randomly diluted
Ising ferromagnet~\cite{cm}.
The latter model has a Yang-Lee edge singularity~\cite{yang} at the same value,
$h=h_c^0$ as
the non-diluted Ising ferromagnet~\cite{griffiths}. Note however that
$h_c^0 = 0$
in random Ising ferromagnets with an unbounded probability
distribution for the local randomness. It has been
argued~\cite{cm} that there is critical field, $h=h_c$, such that for
$h_c^0 < h < h_c$, the zeros of the
partition function are analogous to the localized states in the band tail
in Anderson
localization. (The `Griffiths effects'~\cite{griffiths} leading to this region
also have a parallel in the paramagnetic phase of the quantum spin glass.)
The singularity at
$h=h_c$ is then analogous to a mobility edge~\cite{cm}.
It is this singularity at $h=h_c > 0$, called the `pseudo Yang-Lee edge' in
Ref~\cite{cm},
that interests us here.
The field theory for this singularity is~\cite{mef,geoff,cm}:
\begin{equation}
{\cal A}_{YL} = \int d^d x \left\{ \frac{1}{t} \sum_a \left[ i
\frac{r}{\kappa} \phi^{a} (x) +
\frac{1}{2}(\nabla \phi^a )^2 + i \frac{\kappa}{3} (\phi^a (x) )^3 \right]
+  \frac{1}{2t^2}
\sum_{ab}
\phi^a (x)
\phi^{b} (x) \right\},
\end{equation}
where $\phi^a$ is the replicated order parameter for the Ising model.
This field theoretic model was also considered earlier by
Parisi and Sourlas~\cite{parsou}, who argued that for $\kappa$
{\em imaginary\/}, ${\cal A}_{YL}$ describes the statistics of lattice animals.
As we will argue shortly, the perturbative RG equations for ${\cal A}_{YL}$
are given precisely by (\ref{rgeqn}), and a perturbative
fixed point with $\kappa$ imaginary can indeed be obtained in the $8-d$
expansion. However, in this paper we are only interested in the case of
$\kappa$ real, which also describes the `pseudo Yang-Lee edge'~\cite{cm} in
a random
Ising ferromagnet.

Now we discuss the perturbative connection between models defined by ${\cal
A}_{YL}$ and
${\cal A}$. Consider the Feynman graph expansions with the action ${\cal
A}_{YL}$
 for the correlators
\begin{eqnarray}
G_{YL} (x-y)
&=&\frac{1}{n} \sum_a \left\langle\left\langle \phi^a (x) \phi^a (y)
\right\rangle\right\rangle - G^d_{YL} (x-y)
\nonumber \\
G^d_{YL} (x-y) &=&
\frac{1}{n(n-1)} \sum_{a\neq b} \left\langle\left\langle
\phi^a (x) \phi^b (y) \right\rangle\right\rangle.
\end{eqnarray}
Compare this with the Feynman graph expansion with the action ${\cal A}$ of
{\em zero
frequency\/} correlators
$G$ and $G^d$ respectively. It is not difficult to show, term by term, that
these two expansions are identical to all orders in $\kappa$ and to leading
order in $t$. The fact that ${\cal A}$ involves a matrix field (two
replica indices) while ${\cal A}_{YL}$ has scalar field (one
replica index) does not affect any of the multiplicity factors associated
with any
graph; to leading order in $t$, all relevant graphs were tree graphs before
averging over the disorder in $r_{\mu}$ that corresponds to the
 $1/t^2$ vertex, and none of these graphs have numerical
factors associated with summation over replica or vector indices.
The equality of these perturbative expansions suggests, but does not establish,
that the perturbatively inaccessible static fixed points of ${\cal A}$ and
${\cal A}_{YL}$ may also be identical: if so, any scaling relations satisfied
by
${\cal A}_{YL}$ should apply also to ${\cal A}$.

The non-random Yang-Lee edge problem has a simple scaling
structure~\cite{mef}---there is a scaling relation
between the exponents $\eta$ and $\nu$ as the order parameter $\phi$
is also the ``thermal operator''. The simplest scaling hypothesis for the
random case is that the identification of $\phi$ as the thermal
operator continues to hold. This gives us the scaling relation
\begin{equation}
\frac{1}{\nu} = \frac{d-\theta + 2 - \eta}{2} .
\label{nu}
\end{equation}
Numerical tests of this scaling relation in the randomly diluted Ising
model would
be quite useful (numerical studies of the full quantum spin glass problem
are expected to be
much more difficult). When combined with (\ref{znu}), (\ref{nu}) leads to a
scaling
relation between $z$, $\theta$, and $\eta$ which is very similar (or
identical if
$z\nu=1$) to one
considered recently by Kirkpatrick and Belitz~\cite{kirkbel} for the metal
insulator
transition.

\section{Conclusions}
This paper has proposed a quantum field theory, defined by the action
${\cal A}$,
for
the low energy properties of metallic spin glasses in the vicinity of a $T=0$
transition between a metallic paramagnet and a metallic spin glass.
The mean field phase diagram of the model as a function of a quantum coupling,
temperature and applied magnetic field was described. The phase transitions and
crossovers in this phase diagram were argued to be characteristic of a
zero temperature, {\em static\/} critical theory containing no dynamic
quantum fluctuations. Quantum effects were shown to be dangerously irrelevant
and controlled by a crossover exponent $-\theta_u =-2$.

Next an attempt was made to extend these results beyond mean field theory,
but found runaway flows to strong coupling for all spatial dimensions below
$d=8$. Nevertheless we used the insight gained from the mean field theory to
propose a set of scaling hypotheses.
We assumed that the true critical theory also contained only
static,
randomness induced fluctuations, and the exponent $\theta_u$ controlling
quantum effects took an unknown positive value. This had some
important observable consequences, similar to those found in the mean field
theory:
\newline
({\em i\/}) The non-linear susceptibility had a weaker singularity at the
critical
point than might have been suggested by the usual scaling arguments. In
particular,
for a large enough $\theta_u$ a non-divergent cusp-like singularity was
also possible. \\
({\em ii\/}) In a simple model of the renormalization group flow equations,
finite $T$ dynamic
response functions scaled as functions of
$\omega / T^{1 +
\theta_u \nu}$, rather than the usual scaling as functions of $ \omega /
 T$. \\
({\em iii\/}) Exponents associated with various crossovers in the vicinity
of the
$T=0$ critical point were modified by $\theta_u$.\\
While these results were directly motivated by our analysis of metallic spin
glasses, it is possible that some of the scaling ideas are more general and
could apply also to insulating Ising and rotor spin glasses and other $T=0$
transitions in random quantum systems.

\acknowledgements
The research was supported by NSF Grants No. DMR-92-24290 and DMR-91-57484.
R.O. wishes to thank the Yale University Physics Department for hospitality
extended to him during his visit and the DFG for support
under grant Op28/4-1.\\
{\em Note added:\/} In a recent preprint~\cite{anirvan}, Sengupta and
Georges have considered a model closely related to ours, and obtained results
in general agreement with the mean-field theory of Section~\ref{mft}.

\begin{figure}
\caption{Phase diagram of the action ${\cal A}$ (Eqn
(\protect\ref{landau})) as a
function of temperature $T$ and $r$ which measures
the strength of
quantum fluctuations. The full line
is the only phase transition and dashed lines denote crossovers between
different regimes, which are described in the text.
}
\label{phasediag}
\end{figure}
\end{document}